# Combined Spatial and Temporal Blocking for High-Performance Stencil Computation on FPGAs Using OpenCL

Hamid Reza Zohouri, Artur Podobas, Satoshi Matsuoka
Tokyo Institute of Technology, Tokyo, Japan
{zohouri.h.aa@m,podobas.a.aa@m,matsu@is}.titech.ac.jp

## ABSTRACT

Recent developments in High Level Synthesis tools have attracted software programmers to accelerate their high-performance computing applications on FPGAs. Even though it has been shown that FPGAs can compete with GPUs in terms of performance for stencil computation, most previous work achieve this by avoiding spatial blocking and restricting input dimensions relative to FPGA on-chip memory. In this work we create a stencil accelerator using Intel FPGA SDK for OpenCL that achieves high performance without having such restrictions. We combine spatial and temporal blocking to avoid input size restrictions, and employ multiple FPGA-specific optimizations to tackle issues arisen from the added design complexity. Accelerator parameter tuning is guided by our performance model, which we also use to project performance for the upcoming Intel Stratix 10 devices. On an Arria 10 GX 1150 device, our accelerator can reach up to 760 and 375 GFLOP/s of compute performance, for 2D and 3D stencils, respectively, which rivals the performance of a highly-optimized GPU implementation. Furthermore, we estimate that the upcoming Stratix 10 devices can achieve a performance of up to 3.5 TFLOP/s and 1.6 TFLOP/s for 2D and 3D stencil computation, respectively.

## 1 INTRODUCTION

FPGA designs have traditionally been described using Hardware Description Languages (HDLs) such as VHDL and Verilog. These low-level languages require deep understanding of hardware design, which has prevented large-scale adoption of FPGAs in the High Performance Computing (HPC) community. However, with the recent improvements in High Level Synthesis (HLS), especially the OpenCL programming model, new opportunities have opened up for using FPGAs in HPC.

Stencils are an important computation pattern in HPC, used for solving differential equations, weather, seismic and fluid simulations, and convolution neural networks. Many real-world simulations involve very large 3D stencils with dimensions in the order of tens of thousands of cells that are accelerated using world-class supercomputers [3, 21, 24]. Such stencils are generally so large that even when the problem space is spatially distributed over thousands of nodes, the per-node problem size is still hundreds of cells wide in each dimension. Furthermore, the problem size for such simulations is increasing due to the need for higher resolution and accuracy [6, 12].

Previous work [1, 9, 20, 22] have shown that FPGAs can achieve GPU-level performance in stencil computation. Most of such work achieve this level of performance by relying on temporal blocking *without* spatial blocking. By avoiding spatial blocking, design complexity is significantly reduced and performance can scale near-linearly with the degree of temporal parallelism. However, depending on on-chip memory size, lack of spatial blocking comes at the cost of limiting width for 2D stencils to a few thousands cells [9, 20, 22], and plane size for 3D stencils to 128 × 128 cells or even less [20, 22]. Furthermore, lack of spatial blocking prevents supporting larger input sizes by spatial distribution over multiple FPGAs. Hence, even though such implementations could show the potential of FPGAs for stencil computation, they have limited use cases in accelerating real-world HPC applications. The main motivation of our work is to avoid restricting input size by combing spatial and temporal blocking in a deep-pipelined FPGA design, and show that it is still possible to achieve comparable performance to high-end GPUs. This paves the way for using FPGAs to accelerate real-world stencil-based computations.

We propose a parametrized accelerator based on Intel FPGA SDK for OpenCL that, to the best of our knowledge, achieves the highest performance for stencil computation on a single FPGA without restricting input size. We show that despite limited external memory bandwidth on current FPGA boards, shift register-based spatial blocking coupled with deep-pipelined temporal blocking allows us to achieve comparable performance to high-end GPUs. Our contributions can be summarized as follows:

- We propose a parameterized OpenCL-based FPGA accelerator which combines spatial and temporal blocking. We tackle issues arisen from the added design complexity by utilizing multiple FPGA-specific optimizations, and achieve high throughput without restricting input size.

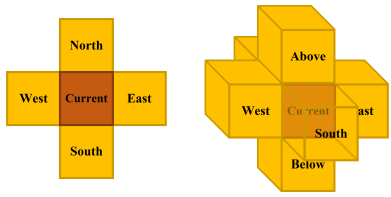

Figure 1: Example of 2D and 3D stencils

- We devise a performance model for predicting performance and pruning parameter search space.
- We evaluate our accelerator with two 2D, and two 3D stencils, each having different memory and compute characteristics. We show that the trade-off between vectorization and degree of temporal parallelism should be exploited in different ways for 2D and 3D stencils, to achieve the best performance.
- We demonstrate that our FPGA-based accelerator can compete with high-end GPUs in terms of performance, and based on a conservative performance projection, estimate that the upcoming Stratix 10 FPGAs can offer similar or even better performance compared to their same-generation GPUs.

## 2 BACKGROUND

### 2.1 Stencil Computation

In stencil computation, a grid (typically 2D or 3D) is iteratively traversed from a starting cell, and each grid cell is updated based on a set of coefficients and the values of its neighboring cells. The stencil radius and shape determine how many cells, and in which directions, are used in the computation. Fig. 1 shows typical first-order (radius of one) 2D 5-point and 3D 7-point stencils. In each iteration (time-step), input cells are read from one buffer, and updated cells are written to another. These two buffers are then swapped before the next iteration is started.

Stencil computation exhibits good spatial locality. To reduce the number of high-latency accesses to external memory and improve performance, the input grid is generally divided into blocks that are read into internal memory. Then, the computation uses this internal buffer instead. This widely-used technique is called *spatial blocking*. Stencil computation also exhibits good temporal locality. When one spatial block is computed and its output is stored in internal memory, it is possible to start computing the next iteration for this block, without fully computing the first iteration for the whole grid. This technique, called *temporal blocking*, allows further reduction of accesses to external memory.

### 2.2 OpenCL

OpenCL is an open, portable standard defined by Khronos Group [10] for writing software targeting heterogeneous platforms in a host/device-based fashion. Using OpenCL involves creating a *host code*, usually written in C/C++, that runs on the host processor, and a *kernel code* written in C, that runs on the device/accelerator. Usually, the host code is compiled using a C/C++ compiler, and the kernel code is compiled at runtime using a compiler that supports the target device. OpenCL provides APIs to manage the device and abstract away the communications between the host and the device. The unit of computation in OpenCL is a *work-item*. Work-items are grouped into *work-groups* in a multidimensional descriptor called an *NDRange*. The work-items in each work-group can share data using *local memory*, which is generally implemented on-chip.

### 2.3 Intel FPGA SDK for OpenCL

Altera SDK for OpenCL (now called Intel FPGA SDK for OpenCL) was released in 2013 [4]. With this SDK, Altera (now Intel PSG) FPGAs became more widely available to software programmers since it allowed them to program FPGAs using a software programming language and a standard API. *Compiling* the kernel code for FPGAs involves compiling OpenCL to LLVM Intermediate Representation (IR) and then to Verilog, followed by the standard EDA flow. Because of this, the kernel code has to be compiled offline, and loaded at run time to reconfigure the FPGA.

Apart from the standard NDRange kernel programming model, Intel FPGA SDK for OpenCL also provides a *single work-item* kernel programming model. In this model, no thread-level parallelism exists, and the compiler will instead extract pipeline-parallelism from the loops in the kernel code. This model more closely matches the traditional deep-pipeline approach of programming FPGAs.

## 3 IMPLEMENTATION

Our design goals are to enable unrestricted input sizes for stencil computation, without sacrificing performance. The first goal requires using spatial blocking, while the low external memory bandwidth of current FPGA boards mandates utilizing temporal blocking to achieve the second one. Combing spatial and temporal blocking creates new challenges, including area overhead and lowered operating frequency due to multiply-nested loops, and memory access alignment issues. Furthermore, our design needs to be parameterized so that we can efficiently use the FPGA area by tuning these parameters. To realize all of these goals, apart from spatial and temporal blocking, we employ multiple FPGA-specific optimizations in a parameterized deep-pipelined OpenCL design.

The outline of our implementation is similar to [16] which targets CPUs and GPUs. We use the single work-item kernel programming model for two reasons. First, shift registers which are the most efficient on-chip storage type for stencil computation can only be inferred in this model. Second, an NDRange implementation will require barrier-based synchronization between threads which will result in pipeline flushes on the FPGA and reduce pipeline efficiency. Hence, we believe that a thread-based implementation using the NDRange model would likely not be able to fully exploit the advantages of FPGAs for stencil computation.

We use a multi-kernel design, with a *read*, a *compute* and a *write* kernel. As shown in Fig. 2, the compute kernel consists of multiple replicated Processing Elements (PEs). Data is streamed from external memory through the PEs using on-chip channels, and is written back to external memory in the end.

### 3.1 Spatial Blocking on FPGAs

We employ spatial blocking to avoid input size restrictions. We use shift registers as on-chip buffers to take advantage of the regular memory access pattern in stencil computation. This is a well-known optimization that is employed in many deep-pipeline implementations of stencil computation on FPGAs [9, 20, 22]. This optimization is not applicable to CPUs and GPUs due to lack of hardware support for this storage type. Furthermore, shift registers cannot be used in a thread-based implementation since they require sequential static addressing known at compile-time.

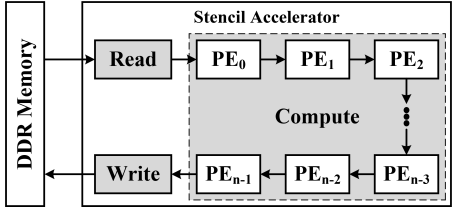

Figure 2: Overview of our multi-kernel design

Fig. 3 shows how all neighbors for a 2-D 5-point stencil are buffered in a shift register. Incrementing the starting address of the buffer shifts the stencil forward, while all the neighbors stay at the same distance relative to the starting point (static addressing). New cells are written to the head of the shift register every clock, and old ones are evicted from the tail. This type of storage can be efficiently mapped to FPGA on-chip Block RAMs.

Using shift registers allows us to minimize the size of on-chip memory buffers by only storing cells of the spatial block that are needed. This is in contrast with spatial blocking on GPUs (or similar highly-threaded hardware) where all of the spatial block is stored on-chip until it has been computed. Supporting shift registers is one of the architectural advantages of FPGAs for stencil computation, which enables us to use larger spatial blocks or more temporal parallelism on FPGAs, compared to GPUs.

We use 1D and 2D spatial blocking for 2D and 3D stencils, respectively. Computation starts from the left or top left block, and each block is computed in all dimensions before going to the next one. Computation of spatial blocks is streamed (no blocking) in the $y$ dimension for 2D, and $z$ dimension for 3D stencils. We also vectorize the computation and coalesce memory accesses simultaneously by loop unrolling. In theory, if the dimensions of the spatial block are $bsize_{\{x|y\}}$, for a stencil of radius $rad$ and a vector size of $par_{vec}$, the size of the shift register will be equal to:

$$size = \begin{cases} 2 \times rad \times bsize_x + par_{vec}, & 2D \\ 2 \times rad \times bsize_x \times bsize_y + par_{vec}, & 3D \end{cases} \quad (1)$$

In practice, due to multiple accesses to the shift register per loop iteration, and limited number of ports per FPGA Block RAM, all or parts of the shift register need to be replicated to support all the parallel accesses. This further increases Block RAM utilization. Altera (Intel) OpenCL Compiler (AOC) automatically performs this operation, while minimizing Block RAM utilization. As we manually cache data, we disable the private cache that is created by AOC for every external memory access to save Block RAMs.

As shown in Fig. 4, we use *overlapped blocking (tiling)* to avoid synchronization between adjacent spatial blocks. Overlapping blocks adds redundant memory accesses and computations, but removes the read-after-write (RAW) dependency between time-steps, allowing us to compute multiple iterations for the same spatial block in parallel. The overlapped parts of the blocks are called *halos* or *ghost zones*. In absence of temporal blocking, the width and height of this region is directly proportional to $rad$, and $bsize_{\{x|y\}}$, respectively, with 2 and 4 such regions existing per spatial block for 2D and 3D stencils, respectively. It is worth noting that we do not require the input dimensions to be divisible by $bsize_{\{x|y\}}$ and hence, there can be a significant amount of out-of-bound computation in the last row and last column of blocks, as also shown in Fig. 4. For inputs that are very large compared to the spatial block size, this overhead will be negligible.

### 3.2 Temporal Blocking on FPGAs

To realize temporal blocking, we use a compute kernel that consists of multiple replicated PEs. Each PE will compute a different time-step of the same spatial block, with a distance of $rad$ rows (for 2D) or planes (for 3D) from the previous PE. Data is transfered between PEs using shallow on-chip channels.

In a standard OpenCL design, having multiple PEs will require creating one kernel for each, and creating multiple queues in the host code to invoke each of the kernels separately. To avoid this issue, we use the *autorun* kernel type provided by Intel FPGA SDK for OpenCL. This kernel type allows any kernel that does not have an interface to host or device memory to be replicated without needing to modify the host code. Each replica in this case can be customized using a static compiler-supplied ID, and will run automatically without needing to be invoked from the host. Another important advantage of this kernel type is that the compiler can better optimize the pipeline and hence, operating frequency scales very well even with tens of PEs in the design.

To use temporal blocking with overlapped spatial blocking, we need to further increase the width of halo regions. In this case, the width of each halo region in the last PE will be:

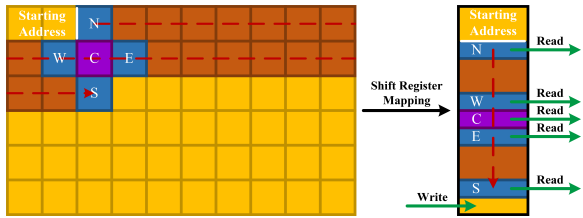

Figure 3: Shift register-based spatial blocking

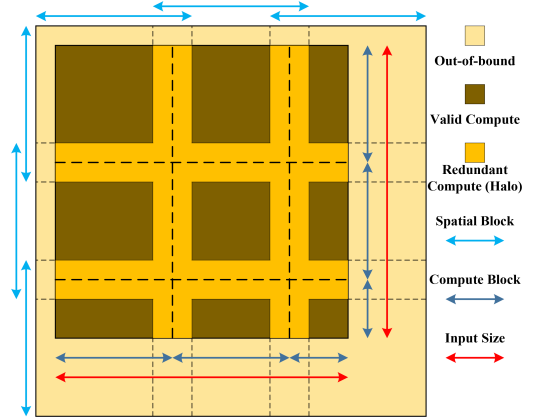

Figure 4: Overlapped blocking (tiling)

$$size_{halo} = rad \times par_{time} \qquad (2)$$

$par_{time}$ is the number of parallel time-steps, which also equals the number of PEs. Fig. 5 shows how the halo size increases as we go further towards the last parallel time-step. Here, block size stays the same regardless of the time-step, and only the region with valid computation will become smaller. Large halos cause *thread divergence* on architectures like GPUs, since the threads that process the halos go through a different path compared to the ones that perform valid computation. Avoiding this issue on GPUs requires complex optimizations like Warp Specialization [14]. In a deep-pipelined FPGA design, however, both paths of a control flow statement are created, where the result of the control flow is multiplexed out based on the evaluated condition. This technique removes flow divergence at the cost of an area penalty. In our design, we reduce this area penalty by redundantly computing halo regions, and only controlling the flow of writes to external memory. Lack of thread divergence and the need for Warp Specialization is another advantage of using FPGAs for stencil computation, which allows better scaling with temporal blocking, compared to GPUs.

When the number of iterations is not a multiple of $par_{time}$, the unused PEs will just forward the data to the next PE in the chain. Even though forwarding adds overhead, as the number of iterations increases, this overhead will diminish.

## 3.3 FPGA-Specific Optimizations

*3.3.1 Loop Collapsing.* Stencil computation with spatial blocking requires multiple nested loops to iterate over dimensions and blocks. Using nested loops in an FPGA design has two disadvantages. First, to achieve an iteration interval (II) of one for all of the loops, the exit conditions of all of them need to be determined in one clock cycle. This creates a long critical path and reduces operating frequency. Second, preserving the state of variables in such loops incurs additional area and memory overhead. Because of these reasons, we collapse all of our loops into one as shown in the conversion from Listing 1 to Listing 2.

**Listing 1: Original**
```
1 for(y = 0; y < m; y++)
2 {
3   for(x = 0; x < n; x++)
4   {
5     compute(x,y);
6   }
7 }
```

**Listing 2: Loop Collapsed**
```
1 int x = 0, y = 0;
2 while(y != m) {
3   compute(x,y);
4   x++;
5   if (x == n) {
6     x = 0;
7     y++; }}
```

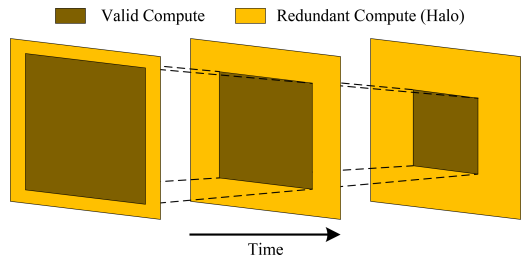

Figure 5: Halo size variance in temporal blocking

*3.3.2 Exit Condition Optimization.* While loop collapsing reduces area usage, the critical path of the design remains unchanged. This critical path is on the loop's exit condition calculation, which consists of a long chain of comparisons and state updates on dimension variables. We shorten this critical path by manually calculating the number of times the collapsed loop will iterate, on the host, and converting the exit condition to a single accumulation followed by an integer comparison. Listing 3 shows how this optimization is applied to Listing 2. This optimization allowed us to increase operating frequency from 200 MHz to over 300 MHz. Our results indicate that after this optimization, the critical path still consists of the remaining comparison and state updates for the dimension variables.

**Listing 3: Exit Condition Optimization**
```
1 int x = 0, y = 0, index = 0;
2 while (index != m * n) {
3   index++;
4   compute(x,y);
5   x++;
6   if (x == n) {
7     x = 0;
8     y++; }}
```

*3.3.3 Padding.* We observed that external memory accesses which are not 512-bit aligned are split by the memory controller at runtime, resulting in significant memory bandwidth waste. In our implementation, valid memory accesses start from the beginning of the top left compute block (brown area in Fig. 4), which is $size_{halo}$ floats apart from the beginning of the spatial block; hence, unless $size_{halo}$ is a multiple of 512 bits, the starting access and every access after that will not be 512-bit aligned. Furthermore, since our spatial blocks overlap, even if the starting point of the first compute block is aligned and $bsize_{\{x|y\}}$ are multiples of 512 bits, the starting point of other blocks might not be 512-bit aligned. Here, the distance between the starting point of two adjacent spatial blocks is equal to $bsize_{\{x|y\}} - 2 \times size_{halo}$. For single-precision floating-point grid cells, assuming that $par_{time}$ is a multiple of eight, and $bsize_{\{x|y\}}$ and input dimensions are divisible by 512 bits, this distance and $size_{halo}$ will be multiples of 512 bits. Because of this, all accesses can be aligned for such values of $par_{time}$. For other cases, however, either $size_{halo}$ or the distance between two adjacent spatial blocks will not be a multiple of 512 bits and hence, some memory accesses could be unaligned.

To alleviate this issue, we pad the device buffers by $par_{time}$ % 8 words. This forces the starting point of the first compute block to be always 512-bit aligned and hence, we can achieve fully-aligned accesses also for values of $par_{time}$ that are multiples of four, and improve performance by over 30%. For other values of $par_{time}$, the distance between two adjacent spatial blocks will not be a multiple of 512 bits and even though alignment will be improved, many accesses will still be unaligned.

## 4 PERFORMANCE MODEL

We created a performance model to predict the impact of different parameters on performance, and perform design space exploration for our accelerator. Table 1 shows the description of parameters we use in our model. Iterative stencil computation generally has higher bytes-to-FLOP ratio compared to what is available on most hardware and hence, is generally memory-bound [16]. For our model we assume the computation is memory-bound, and external memory latency is hidden by the deep pipeline. To predict run time and throughput, we need to estimate external memory throughput and correctly count the number of accesses to external memory.

Table 1: Parameters Description

| Parameter | Description | Unit |
|---|---|---|
| $rad$ | Stencil radius | Cells |
| $par_{vec}$ | Compute vector size (width) | N/A |
| $par_{time}$ | Number of parallel time-steps | N/A |
| $f_{max}$ | Kernel operating frequency | Hz |
| $size_{cell}$ | Size of each grid cell | Bytes |
| $size_{input}$ | Number of cells in input grid | Cells |
| $size_{halo}$ | Width of each halo region | Cells |
| $num_{read}$ | External memory reads per cell update | N/A |
| $num_{write}$ | External memory writes per cell update | N/A |
| $num_{acc}$ | External memory accesses per cell update | N/A |
| $bsize_{\{x|y\}}$ | Size of spatial block in $x$ or $y$ dimension | Cells |
| $csize_{\{x|y\}}$ | Size of compute block in $x$ or $y$ dimension | Cells |
| $th_{mem}$ | Memory throughput | GB/s |
| $th_{max}$ | Maximum memory throughput | GB/s |
| $dim_{\{x|y|z\}}$ | Input size per dimension | Cells |
| $bnum_{\{x|y\}}$ | Number of spatial blocks per dimension | Cells |
| $trav_{\{x|y\}}$ | Number of traversed cells per dimension | Cells |
| $iter$ | Number of iterations | N/A |

We observe that external memory performance on FPGAs scales with both kernel operating frequency, $f_{max}$, and compute vector size, $par_{vec}$, until the maximum throughput of the external memory is reached. This maximum throughput is determined by the memory bus width and the frequency of external memory DIMMs. We estimate memory performance in GB/s[1] as follows:

$$num_{acc} = num_{read} + num_{write}$$
$$th_{mem} = min\left(\frac{f_{max} \times par_{vec} \times size_{cell} \times num_{acc}}{10^9}, th_{max}\right) \quad (3)$$

To calculate the number of memory accesses we first calculate the number of total accesses to external memory, including the redundant and out-of-bound ones. We define *compute block* as the region in each spatial block with only valid computation (dark blue arrows in Fig. 4). The dimensions of the compute block are:

$$csize_{\{x|y\}} = bsize_{\{x|y\}} - 2 \times size_{halo} \quad (4)$$

Since the spatial blocks are overlapped in a way that in the last PE, the compute blocks are consecutive (as seen in Fig. 4), each dimension of the input is traversed up to a point that the index in that dimension is a multiple of $csize_{\{x|y\}}$. Hence, number of spatial/compute blocks in each dimension is:

$$bnum_{\{x|y\}} = \left\lceil \frac{dim_{\{x|y\}}}{csize_{\{x|y\}}} \right\rceil \quad (5)$$

---
[1]All throughput numbers in this paper are in GB/s = $10^9$ B/s, and not GiB/s = $2^{30}$ B/s

Consequently, the number of cells that are read from external memory for each input buffer is calculated as follows:

$$t_{cell} = \begin{cases} bnum_x \times bsize_x \times dim_y, & 2D \\ bnum_x \times bsize_x \times bnum_y \times bsize_y \times dim_z, & 3D \end{cases} \quad (6)$$

Since we avoid out-of-bound memory reads and writes, and also memory writes to halo regions, the number of reads from external memory will be equal to $t_{cell}$ minus out-of-bound cells, multiplied by the number of reads per cell update. The number of writes will also be equal to input size multiplied by the number of writes per cell update. For example, the total number of reads from external memory for a 2D stencil will be:

$$trav_x = bnum_x \times csize_x + 2 \times size_{halo}$$
$$t_{read} = \left(t_{cell} - (trav_x - dim_x) \times dim_y\right) \times num_{read} \quad (7)$$

Now we can calculate run time (seconds) and throughput (GB/s):

$$run\_time = \frac{\left\lceil \frac{iter}{par_{time}} \right\rceil \times (t_{read} + t_{write}) \times size_{cell}}{10^9 \times th_{mem}} \quad (8)$$

$$throughput = \frac{num_{acc} \times size_{input} \times size_{cell} \times iter}{10^9 \times run\_time} \quad (9)$$

Throughput can be converted to compute performance (GFLOP/s) by using the bytes-to-FLOP ratio of the stencil.

## 5 METHODOLOGY
### 5.1 Benchmarks

For evaluating our accelerator, we use four stencils, two 2D and two 3D. We use the 2D and 3D version of the Hotspot benchmark from Rodinia Benchmark Suite [2], and also Diffusion 2D and 3D [14]. Table 2 shows the computation and characteristics of these stencils. The bytes per cell update (Bytes PCU) numbers reported in this table assume full spatial locality optimization.

All of these benchmarks use single-precision floating-point numbers. All the variables except $TEMP_{AMB}$ (compile-time constant) are passed to the kernel as arguments in form of values or arrays, and can be changed without kernel recompilation. The subscripts show direction (current, north, south, west, east, above, and below). Apart from arithmetic intensity, Hotspot also differs from Diffusion in memory characteristics since it needs two reads from external memory per cell update. In our implementation, both reads are cached using shift registers, though the shift register for the *power* input will be smaller than the one used for the main input since only the *current* value needs to be cached. In all of the stencils, all out-of-bound neighbors of grid cells on the grid boundaries will fall back on the boundary cell itself.

Table 2: Benchmarks

| Benchmark | Computation | FLOP PCU | Bytes PCU | $\frac{Bytes}{FLOP}$ |
|---|---|---|---|---|
| Diffusion 2D | $c_c \times val_c + c_w \times val_w + c_e \times val_e + c_s \times val_s + c_n \times val_n$ | 9 | 8 | 0.889 |
| Diffusion 3D | $c_c \times val_c + c_w \times val_w + c_e \times val_e + c_s \times val_s + c_n \times val_n + c_b \times val_b + c_a \times val_a$ | 13 | 8 | 0.615 |
| Hotspot 2D | $val_c + sdc \times (power_c + (val_n + val_s - 2.0 \times val_c) \times Ry\_1 + (val_e + val_w - 2.0 \times val_c) \times Rx\_1 + (TEMP_{AMB} - val_c) \times Rz\_1)$ | 15 | 12 | 0.800 |
| Hotspot 3D | $val_c \times c_c + val_n \times c_n + val_s \times c_s + val_e \times c_e + val_w \times c_w + val_a \times c_a + val_b \times c_b + sdc \times power_c + c_a \times TEMP_{AMB}$ | 17 | 12 | 0.706 |

## 5.2 Hardware and Software Setup

We evaluate our implementation on the Terasic DE5-net board with a Stratix V GX A7 FPGA, and the Nallatech 385A board with an Arria 10 GX 1150 FPGA. We compare our results with four generations of high-end NVIDIA GPUs. Table 3 contains a comprehensive comparison of these devices. To keep comparison fair, we disable ECC on the GPUs. We compile our OpenCL host code using GCC 5.3.1, and kernel code using Quartus and AOC v16.1.2, and use CUDA v8.0/9.0 for compiling GPU kernels.

For power measurement, we use the NVIDIA NVML [18] library for GPUs, and Nallatech's API on the Arria 10 board, to access the on-board power sensors. In both cases the sensor is queried every 10 ms during kernel execution, and average power is calculated. For the Stratix V board, no power sensor exists on the board. As a conservative estimate, we run Quartus PowerPlay on the place-and-routed OpenCL design with a 25% toggle rate (default is 12.5%) and add 2.34 Watts (obtained from the datasheet of a similar memory model [11]) to the obtained value as the maximum power consumption of the external memory.

We choose $dim_{\{x|y\}}$ to be a multiple of $csize_{\{x|y\}}$ to minimize the number of out-of-bound cells, and clearly show the potential of our accelerator. We use square and cubic inputs, for 2D and 3D stencils, respectively, with at least 1 GB of external memory usage. We observe that as long as $dim_{\{x|y\}}$ are multiples of $csize_{\{x|y\}}$, performance variation with input size is negligible. We only measure kernel run time, and ignore initialization and data transfer time between host and device. Each benchmark is run with 1000 iterations, and average of five runs is reported. In our case, benchmark run times were at least 3 seconds for 2D, and 7 seconds for the 3D stencils, with a variation of less than 5 ms.

## 5.3 Parameter Tuning

To achieve maximum performance with the available FPGA area, we need to tune three parameters: $bsize_{\{x|y\}}$, $par_{vec}$ and $par_{time}$. Increasing $bsize_{\{x|y\}}$ reduces redundancy and improves performance scaling with higher $par_{time}$. However, both of these values affect Block RAM utilization which creates an area/performance trade-off. Another such trade-off exists between $par_{vec}$ and $par_{time}$, both of which increase performance (with different scaling factors) at the cost of higher DSP utilization. We use the area report generated by AOC to determine how many DSPs are necessary for one cell update, and then use our model and the device DSP count to optimize the trade-off between $par_{time}$ and $par_{vec}$. Predicting Block RAM utilization is not straightforward due to Block RAM packing during mapping, and also mapping of some buffers to distributed memory instead of Block RAMs. Hence, we experimentally optimize the trade-off between $par_{time}$ and $bsize_{\{x|y\}}$. Furthermore, we put the following restrictions on our parameters:

- We use square spatial blocks for 3D stencils. Even though our implementation allows non-square blocks, doing so reduces our parameter search space with little effect on performance.
- We assume $bsize_{\{x|y\}}$ are powers of two so that the block indexes can be updated using an efficient *mod* operation. Other block sizes can be supported using conditional branching, at the cost of 5-20 MHz lower operating frequency.
- $bsize_x$ must be divisible by $par_{vec}$
- Since the compiler only creates coalesced access ports to external memory with a width that is a power of two, we limit $par_{vec}$ to powers of two to avoid bandwidth waste.
- We prefer multiples of four for $par_{time}$ to avoid unaligned accesses.

Performance predictions from our model combined with the compiler's area report allow us to limit the number of candidate configurations per stencil per board to less than six, which significantly reduces the time and compute resources that are needed for placement and routing.

## 5.4 Compiler Optimizations

*5.4.1 Flat Compilation.* Using Intel FPGA SDK for OpenCL, FPGA reconfiguration is automatically performed at runtime by the OpenCL runtime. On Stratix V devices, this reconfiguration is performed using Configuration via Protocol (CvP). However, CvP update is not supported on Arria 10 [8] and runtime reconfiguration instead happens using Partial Reconfiguration (PR) through PCI-E. Due to the additional placement and timing constraints imposed on placement and routing to support PR, fitting and routing quality for OpenCL kernels is reduced on Arria 10, especially when area utilization is high. Because of this, we used flat compilation for this device which disables PR and place and routes the OpenCL kernel and the Board Support Package (BSP) as a flat design. In our experience, most of our best-performing kernels either failed to fit or route with the default PR-based flow, or exhibited noticeably lower operating frequency compared to the flat flow (up to 100 MHz lower). However, flat runtime reconfiguration happens through JTAG with takes longer (15-20 seconds vs. less than 5 seconds for PR through PCI-E).

*5.4.2 Seed and $F_{max}$ Sweep.* By default, AOC balances the pipeline stages for a target $f_{max}$ of 240 MHz. It is possible to increase this value to achieve higher operating frequency, at the cost of extra logic and memory utilization. For each stencil on each device, we compile all candidate configurations using the default $f_{max}$ target and measure their performance on the device. Then, to eliminate the effect of $f_{max}$ variability, we normalize the measured values for a fixed $f_{max}$ to find the best-performing candidate. After that we recompile the best version with multiple $f_{max}$ targets higher than default, as long as II remains one, to maximize its $f_{max}$. If logic utilization is high (>80%), increasing $f_{max}$ target will instead reduce $f_{max}$ due to more routing congestion. In such cases we instead change the random seed for placement and routing to maximize $f_{max}$.

Table 3: Hardware Comparison

| Device | Peak Memory Bandwidth (GB/s) | Peak Compute Performance (GFLOP/s) | Production Node (nm) | Transistors (Billion) | On-chip Memory (MiB) [a] | On-board Memory (GiB) | TDP (Watt) | Release Year |
|---|---|---|---|---|---|---|---|---|
| Stratix V GX A7 | 25.6 | 200 | 28 | 3.8 | 6.25 + 0.895 | 4 | 40 | 2011 |
| Arria 10 GX 1150 | 34.1 | 1450 | 20 | 5.3 | 6.62 + 1.585 | 8 | 70 | 2014 |
| Tesla K40c | 288.4 | 4300 | 28 | 7.08 | 3.75 + 1.5 | 12 | 235 | 2013 |
| GTX 980Ti | 336.6 | 6900 | 28 | 8 | 5.5 + 3 | 6 | 275 | 2015 |
| Tesla P100 PCI-E | 720.9 | 9300 | 16 | 15.3 | 14 + 4 | 16 | 250 | 2016 |
| Tesla V100 SXM2 | 900.1 | 14900 | 12 | 21.1 | 20 + 6 | 16 | 300 | 2017 |

[a]FPGAs: M20K + MLAB, GPUs: Register + L2

## 6 RESULTS

### 6.1 FPGA Performance

Table 4 shows the results for all of our evaluated stencils on both FPGA boards. The highest estimated performance (adjusted to post-place-and-route $f_{max}$ for correct accuracy calculation) for each kernel on each board is marked in yellow, the highest measured performance is marked in green, and the resource bottleneck for the best configuration on each board is marked in red.

We achieve over twice higher throughput in 2D stencils, versus 3D. This is expected since the much higher Block RAM requirement of 3D stencils significantly reduces $bsize_{\{x|y\}}$, and limits scaling with temporal parallelism. For 2D stencils, however, since $bsize_x$ is sufficiently-high, redundancy is minimized and we can achieve close-to-linear scaling with temporal parallelism. This difference brings us to a very important conclusion: **For 3D stencils, it is better to spend FPGA resources to support a larger vector size, rather than more temporal parallelism**, since the former allows better performance scaling. **For 2D stencils**, however, **it is more efficient to spend FPGA resources on increasing temporal parallelism, rather than vector size**; the latter achieves close-to-linear performance scaling, while performance scaling with the former depends on the behavior of the memory controller which in our experience, is sub-linear except for very small vector sizes (up to four). Still, higher degree of temporal parallelism will result in higher logic utilization and consequently, more routing complications and lower $f_{max}$. Because of this, using the highest $par_{time}$ and lowest $par_{vec}$ will not necessarily result in the highest performance.

For the 2D stencils on Stratix V, Hotspot achieves higher throughput than Diffusion despite lower $par_{time}$. This is due to the fact that the higher $num_{acc}$ in Hotspot allows better utilization of the memory bandwidth with the narrow vector size. It is not possible to fully utilize the DSPs on Stratix V for Hotspot since this stencil has a high number of floating-point addition and subtractions which are not natively supported by the DSPs on this device and hence, performance scaling is constrained by logic utilization. On Arria 10, however, throughput is 40% higher in Diffusion compared to Hotspot since both are constrained by DSP utilization, while the much lower compute intensity of Diffusion allows a twice wider vector at the same $par_{time}$. This is enough to offset the better memory bandwidth utilization of Hotspot due to higher $num_{acc}$. This 40% difference is exactly equal to the ratio of $num_{acc} \times par_{vec} \times f_{max}$ between these two stencils.

For the 3D stencils on Stratix V, total degree of parallelism ($par_{vec} \times par_{time}$) is the same and computation throughput is very close. Hotspot 3D achieves lower $f_{max}$ due to 100% Block RAM and DSP utilization, but this is offset by the higher $num_{acc}$ in this stencil. Also on Arria 10, the computation throughput of the 3D stencils is close. On this device, Diffusion 3D benefits from the higher total degree of parallelism and bigger $bsize_{\{x|y\}}$, while Hotspot 3D benefits from higher $num_{acc}$ and $f_{max}$.

As shown in Table 4, we achieve an $f_{max}$ of over 300 MHz in cases that routing is not constrained by area utilization. This shows that our implementation maps well to the underlying FPGA architecture, and that we have been successful in optimizing the critical path. Since 2D stencils have less dimension variables, their critical path is shorter compared to 3D stencils, and $f_{max}$ is higher.

As a final note on power consumption, in many cases we are using over 70 Watts on the Arria 10 board, which is over its TDP. This further asserts that we are pushing the boundaries of performance on this device.

### 6.2 Model Accuracy

We define model accuracy as ratio of the measured performance on the board, to estimated performance by our model for a fixed $f_{max}$. As show in Table 4, even though our model can correctly predict the trend of performance for different configurations, for 2D stencils we achieve 65-90% of the estimated performance and for 3D we achieve 55-70%. One reason for this discrepancy is that even though we assume memory performance scales linearly with $f_{max}$ and $par_{vec}$, in practice, scaling with $par_{vec}$ is sub-linear except for very small values. Scaling with $f_{max}$ also depends on the effectiveness of the memory controller in runtime coalescing. Linear scaling can be achieved if the kernel $f_{max}$ is lower than the operating frequency of the memory controller (200 and 266 MHz for Stratix V and Arria 10, respectively), but this linearity is lost for higher values which is the general case in our implementation. Apart from this, since more data is read from external memory, than written to it, writes are more likely to be stalled and such stalls can potentially propagate all the way to the top of the pipeline. Since halo regions are not written to external memory, some writes need to be masked and potentially split into two or more accesses by the memory controller. This further increases bandwidth waste and lowers performance and model accuracy. Profiling the kernels using Intel's OpenCL profiler shows that the average burst size is always lower than $par_{vec}$, and does not go beyond eight words, which implies some accesses are being split into smaller ones at runtime.

Table 4: FPGA Results

| Kernel | Device | bsize | $par_{vec}$ | $par_{time}$ | dim | Estimated Performance (GB/s) | Measured Performance (GB/s\|GFLOP/s\|GCell/s) | $f_{max}$ (MHz) | Logic | Memory (Bits\|Blocks) | DSP | Power (Watt) | Model Accuracy |
|---|---|---|---|---|---|---|---|---|---|---|---|---|---|
| Diffusion 2D | S-V | 4096 | 8 | 6 | 16336 | 107.861 | 93.321\|104.986\|11.665 | 281.76 | 62% | 10%\|32% | 95% | 26.575 | 86.5% |
| | | 4096 | 4 | 12 | 16288 | 111.829 | 97.440\|109.620\|12.180 | 294.20 | 63% | 14%\|40% | 95% | 27.509 | 87.1% |
| | | 4096 | 2 | 24 | 16192 | 114.720 | 99.582\|112.030\|12.448 | 302.48 | 69% | 22%\|52% | 95% | 29.845 | 86.8% |
| | A-10 | 4096 | 16 | 16 | 16256 | 540.119 | 359.664\|404.622\|44.958 | 311.62 | 46% | 20%\|45% | 85% | 53.447 | 66.6% |
| | | 4096 | 8 | 36 | 16096 | 780.500 | 673.959\|758.204\|84.245 | 343.76 | 55% | 38%\|83% | 95% | 72.530 | 86.3% |
| | | 4096 | 4 | 72 | 15808 | 635.003 | 542.196\|609.971\|67.775 | 281.61 | 67% | 65%\|100% | 95% | 65.310 | 85.4% |
| Hotspot 2D | S-V | 4096 | 8 | 6 | 16336 | 153.068 | 110.452\|138.065\|9.204 | 272.47 | 91% | 13%\|43% | 77% | 33.654 | 72.2% |
| | | 4096 | 4 | 12 | 16288 | 128.667 | 112.206\|140.258\|9.351 | 225.83 | 95% | 21%\|53% | 77% | 24.271 | 87.2% |
| | | 4096 | 2 | 20 | 16224 | 128.950 | 112.218\|140.273\|9.352 | 269.97 | 84% | 27%\|61% | 64% | 33.361 | 87.0% |
| | A-10 | 4096 | 8 | 16 | 16256 | 468.024 | 355.043\|443.804\|29.587 | 308.35 | 39% | 27%\|42% | 85% | 41.623 | 75.9% |
| | | 4096 | 4 | 36 | 16096 | 547.904 | 474.292\|592.865\|39.524 | 322.47 | 47% | 53%\|94% | 95% | 50.129 | 86.6% |
| | | 4096 | 2 | 72 | 15808 | 483.921 | 415.012\|518.765\|34.584 | 287.43 | 72% | 88%\|100% | 95% | 52.179 | 85.8% |
| Diffusion 3D | S-V | 256 | 8 | 4 | 744 | 75.422 | 62.435\|101.457\|7.804 | 301.02 | 62% | 36%\|67% | 91% | 21.135 | 82.8% |
| | | 256 | 8 | 5 | 738 | 59.019 | 39.918\|64.867\|4.990 | 189.50 | 72% | 44%\|81% | 100% | 22.825 | 67.6% |
| | A-10 | 256 | 16 | 8 | 720 | 261.159 | 178.784\|290.524\|22.348 | 294.81 | 38% | 65%\|76% | 60% | 57.083 | 68.5% |
| | | 256 | 16 | 12 | 696 | 379.230 | 230.568\|374.673\|28.821 | 286.61 | 60% | 94%\|100% | 89% | 71.628 | 60.8% |
| | | 128 | 8 | 24 | 640 | 282.839 | 160.222\|260.361\|20.028 | 308.64 | 52% | 52%\|96% | 89% | 73.208 | 56.6% |
| Hotspot 3D | S-V | 256 | 8 | 4 | 496 | 92.527 | 63.603\|90.104\|5.300 | 246.18 | 76% | 68%\|100% | 100% | 36.126 | 68.7% |
| | | 128 | 4 | 8 | 560 | 78.818 | 61.157\|86.639\|5.096 | 238.32 | 74% | 37%\|76% | 100% | 34.085 | 77.6% |
| | A-10 | 128 | 16 | 8 | 560 | 235.145 | 165.876\|234.991\|13.823 | 256.47 | 45% | 37%\|73% | 77% | 53.933 | 70.5% |
| | | 128 | 8 | 16 | 576 | 321.361 | 194.406\|275.409\|16.201 | 299.85 | 47% | 67%\|100% | 77% | 66.210 | 60.5% |
| | | 128 | 8 | 20 | 528 | 355.284 | 228.149\|323.211\|19.012 | 296.20 | 62% | 81%\|100% | 96% | 73.398 | 64.2% |

Table 5: Stratix 10 Device Specifications

| Device | DSP | Memory Blocks | External Memory Spec. | External Memory Bandwidth (GB/s) |
|---|---|---|---|---|
| GX 2800 | 5,760 (3.8x) | 11,721 (4.3x) | 4-bank DDR4-2400 [15] | 76.8 (2.25x) |
| MX 2100 | 3,744 (2.5x) | 6,501 (2.4x) | 4-tile HBM [13] | 512 (15x) |

For 2D stencils, since $par_{vec}$ is small, this issue has a smaller effect on performance, but for the 3D stencils, the effect is larger since we use higher values of $par_{vec}$. This is the reason why 2D stencils achieve better model accuracy compared to 3D ones.

In Table 4, our model correctly predicts the best configuration in every case, except for Hotspot 2D on Stratix V. The reason is that, as discussed in Section 3.3.3, fully-aligned accesses can only be achieved if $par_{time}$ is a multiple of four and hence, a configuration with $par_{time} = 6$ cannot achieve the predicted performance.

### 6.3 Performance Projection for Stratix 10

To evaluate the potential of future FPGAs for stencil computation, we use our model to predict the performance of two of the upcoming Stratix 10 devices. Table 5 shows the specifications of these devices, and improvement ratio compared to Arria 10 GX 1150.

Designs on the Stratix 10 family are expected to reach an $f_{max}$ of up to 1 GHz, enabled by the latest 14 nm manufacturing node and HyperFlex technology [7]. The extended register insertion and re-timing capabilities offered by HyperFlex are expected to improve $f_{max}$ in case of routing congestion. However, when $f_{max}$ is instead limited by the critical path in the design, HyperFlex will have limited effect. For the specific case of stencil computation, as discussed in Section 3.3.2, the critical path of the design will be the chain of operations that update the state of dimension variables and hence, we expect limited $f_{max}$ improvement with HyperFlex on Stratix 10 devices. Due to this reason, we only assume a conservative 100-MHz increase in $f_{max}$ compared to Arria 10 for stencil computation.

To predict the performance of our stencils for Stratix 10, we estimate the DSP and memory utilization on these devices by extrapolating usage on Arria 10. We assume the devices will have enough logic available to support every configuration. For memory utilization we assume overutilization only if the bits count goes above 100%. Then we use our model alongside with the area utilization estimations to predict the best configuration and performance on Stratix 10. Table 6 shows our estimation results for 5000 iterations and an input size that is a multiple of $csize_{\{x|y\}}$. Based on measured model accuracy from real executions (Table 4), we use a calibration factor of 80% and 60%, for 2D and 3D stencils, respectively, to calibrate our predictions on Stratix 10.

Even with a conservative estimation, we expect the high DSP and Block RAM count of the Stratix 10 GX 2800 device to allow over 3.5 TFLOP/s of compute performance for 2D stencil computation, which will likely outperform its same-generation GPUs. Furthermore, we expect the high memory bandwidth of the MX 2100 device to enable up to 1.6 TFLOP/s for 3D stencils, which will be competitive against its same-generation GPUs. Even though the MX 2100 device has much higher external memory bandwidth, we predict that it will achieve only slightly higher performance compared to GX 2800 for 3D stencils. This is due to the fact that the MX 2100 device has much less resources, and the computation becomes area-bound before the external bandwidth can be fully utilized. **We conclude that a too-high or too-low "external memory bandwidth to compute performance" ratio on FPGAs will result in either area or memory bandwidth bottleneck for stencil computation.**

Table 6: Stratix 10 Performance Estimation

| FPGA | Stencil | bsize | $par_{vec}$ | $par_{time}$ | $f_{max}$ (MHz) | Calibration Factor | Performance (GB/s\|GFLOP/s) | Used Memory Bandwidth (GB/s\|%) | Memory Utilization (Bits\|Blocks) | DSP Utilization |
|---|---|---|---|---|---|---|---|---|---|---|
| GX 2800 | Diffusion 2D | 8192 | 8 | 140 | 450 | 80% | 3162.7\|3558.0 | 28.8\| 38% | 59%\| 88% | 97% |
| GX 2800 | Hotspot 2D | 8192 | 4 | 140 | 450 | 80% | 2362.8\|2953.5 | 21.6\| 28% | 80%\| 91% | 97% |
| GX 2800 | Diffusion 3D | 256 | 32 | 24 | 400 | 60% | 917.4\|1490.8 | 76.8\|100% | 44%\| 47% | 93% |
| GX 2800 | Hotspot 3D | 256 | 16 | 24 | 400 | 60% | 868.8\|1230.8 | 76.8\|100% | 91%\|100% | 61% |
| MX 2100 | Diffusion 2D | 8192 | 8 | 92 | 450 | 80% | 2078.6\|2338.5 | 28.8\| 6% | 69%\|100% | 98% |
| MX 2100 | Hotspot 2D | 8192 | 4 | 92 | 450 | 80% | 1555.0\|1943.8 | 21.6\| 4% | 94%\|100% | 98% |
| MX 2100 | Diffusion 3D | 512 | 128 | 4 | 400 | 60% | 975.3\|1584.8 | 409.6\| 80% | 53%\| 56% | 96% |
| MX 2100 | Hotspot 3D | 256 | 32 | 12 | 400 | 60% | 991.1\|1404.1 | 153.6\| 30% | 81%\|100% | 93% |

### 6.4 Comparison with GPUs

To avoid biased comparisons, we only compare our Diffusion 3D results with the highly-optimized implementation from [14]. We tune the parameters from this implementation for every GPU, and use an input size of $512^3$ which achieves the best performance on these devices. This GPU implementation restricts $dim_{\{x|y\}}$ to values that are a multiple of $bsize_{\{x|y\}}$. Even though Rodinia includes CUDA implementations of both Hotspot 2D and 3D, these implementations are not optimized well, to the point that our implementations on Arria 10 achieve over twice the performance of Tesla P100; hence, we avoid using them for comparison.

Fig. 6 shows the performance and power efficiency of our implementation of Diffusion 3D on FPGAs, compared to GPUs. The roofline performance is the achievable GFLOP/s by full utilization of external memory bandwidth on each device, without temporal blocking. We also add our estimated performance and power efficiency for the Stratix 10 MX 2100 device. Based on [17], we estimate the power consumption of the Stratix 10 GX 2800 FPGA between 140 to 150 Watts for an $f_{max}$ of 400 to 450 MHz. For the smaller MX 2100 device, we will assume a typical power consumption of 125 watts for estimating its power efficiency.

As seen in the graph, we achieve higher performance on Arria 10 compared to Tesla K40c, despite more than eight times lower memory bandwidth. **The performance advantage of Arria 10 is due to better scaling of temporal blocking on FPGAs, compared to GPUs, which allows achieving multiple times higher performance than the roofline.** Despite the fact that Arria 10 cannot reach the performance of the more modern GPUs, our results clearly show the advantage of FPGAs for stencil computation over GPUs. It noteworthy that Arria 10 also achieves better power efficiency compared to GTX 980 Ti, which further asserts the superior power efficiency of FPGAs compared to GPUs of their age. Based on our estimation, the upcoming Stratix 10 MX 2100 FPGA will achieve better performance and power efficiency compared to the Tesla P100 GPU, and better power efficiency compared to the state-of-the-art Tesla V100.

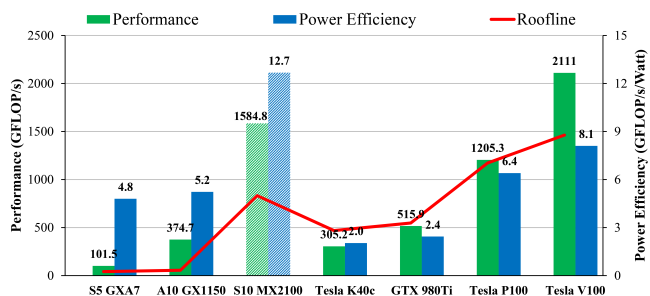

Figure 6: Performance comparison with GPUs

As a final note, we compare the code complexity of our FPGA implementation against the GPU code from [14]. Our FPGA kernel is ~250 lines and can be ported to same-shape stencils in a matter of minutes, and to same-order but differently-shaped stencils in a few hours. Furthermore, block size, vector width and degree of temporal parallelism have been parameterized in our implementation. In contrast, the GPU kernel is ~400 lines, only parameterizes block size, and requires more effort for porting to other stencils.

## 7 RELATED WORK

In [25] we reported an early optimization analysis of Hotspot 2D which only achieved comparable *power efficiency* to GPUs due to lack of temporal blocking and the FPGA-specific optimizations discussed here. In [23], the authors implement multiple stencils using Xilinx SDAccel with both spatial and temporal blocking. Their work uses a thread-based implementation which, as discussed in Section 3, cannot use shift register-based spatial blocking. They also do not employ 3.5D blocking [16]. Since they do not report run time or FLOP/s, we cannot compare our results with theirs. [5, 19] are examples of similar automated frameworks for stencil computation on FPGAs that use dependency analysis and the polyhedral model. These frameworks focus on automation rather than achieving high performance, and use thread-based implementations which suffer from the same shortcoming as [23]. [5] reports 8 GFLOP/s for Jacobi 2D, while we achieve over 110 GFLOP/s on Stratix V (and much more on Arria 10) for Diffusion 2D which has the exact same stencil characteristics. We achieve this large performance advantage despite the fact that the Kintex-7 XC7Z045 FPGA they use has more DSPs and roughly half of the logic and Block RAM count of our Stratix V A7 FPGA.

[1, 9, 20, 22] present the recent high-performing deep-pipelined implementations of stencil computation on FPGAs, all of which avoid spatial blocking and hence, put hard limits on input dimensions relative to on-chip memory size. In contrast, we do employ spatial blocking to avoid such restrictions which limit usability in real-world HPC applications, and show that it is still possible to achieve high performance. Compared to [22], we achieve only 9% lower performance on the same Stratix V device, but with an input size that is not supported by their implementation unless it is modified to use bigger shift registers at the cost of multiple times lower degree of temporal parallelism. In that case, our implementation will have a clear performance advantage. Compared to [9], we achieve 4x higher performance in Hotspot 2D which has similar characteristics to their FDTD 2D (same $num_{acc}$ and one higher FLOP PCU). Compared to [1], we achieve 5x and 40x higher performance for Diffusion 2D and 3D on Stratix V A7, respectively, compared to their results for Jacobi 2D and 3D on a Virtex-7 XC7VX485T FPGA. [20] uses a 2nd order stencil and hence, their results are not comparable with ours.

## 8 CONCLUSION

We studied the potential of FPGAs for accelerating 2D and 3D stencil computation in real-world HPC applications. Using combined spatial and temporal blocking allowed us to, unlike many previous work on FPGAs, achieve high performance without restricting input size. With a parameterized OpenCL-based design and a performance model to guide parameter tuning, we achieved a compute performance of up to 760 and 375 GFLOP/s on an Arria 10 device, for 2D and 3D stencil computation, respectively, which rivals the performance of a highly-optimized implementation on high-end GPUs. Furthermore, we used our performance model for estimating the performance of two of the upcoming Stratix 10 devices in stencil computation, and showed that these devices will be even more competitive against their same-generation GPUs.

Since many real-world HPC applications use high-order stencils, investigating the effectiveness of temporal blocking on FPGAs for such stencils is the subject of our future work. Furthermore, we plan to evaluate spatial distribution of large stencils on multiple FPGAs for accelerating real-world HPC applications in future.

## ACKNOWLEDGMENTS

This work was supported by MEXT, JST-CREST under Grant Number JPMJCR1303, JSPS KAKENHI under Grant Number JP16F16764, the JSPS Postdoctoral fellowship under grant P16764, and performed under the auspices of the Real-world Big-Data Computation Open Innovation Laboratory, Japan. We would like to thank Intel for donating licenses for their FPGA toolchain through their university program, and also thank Mohamed Wahib Attia for helping us with extracting the GPU results.